\documentclass[prb,preprint,showpacs,preprintnumbers,amsmath,amssymb]{revtex4}

\usepackage{graphicx}
\usepackage{dcolumn}
\usepackage{bm}
\usepackage{color}

\def\vk{{\bf k}}

\newcommand{\eq}[1]{Eq.~(\ref{#1})}
\newcommand{\fig}[1]{Fig.~\ref{#1}}
\newcommand{\be}{\begin{equation}}
\newcommand{\ee}{\end{equation}}
\newcommand{\bea}{\begin{eqnarray}}

\newcommand{\eea}{\end{eqnarray}}

\begin{document}   
\title{Electronic Raman scattering from orbital nematic fluctuations}
\author{Hiroyuki Yamase}
\affiliation{Max-Planck-Institut f\"ur Festk\"orperforschung,
             Heisenbergstrasse 1, D-70569 Stuttgart, Germany}
\affiliation{National Institute for Materials Science, Tsukuba 305-0047, Japan}
\author{Roland Zeyher}
\affiliation{Max-Planck-Institut f\"ur Festk\"orperforschung,
             Heisenbergstrasse 1, D-70569 Stuttgart, Germany}

\date{\today}

\begin{abstract}
We compute Raman scattering intensities via the lowest-order coupling to the 
bosonic propagator associated with orbital nematic fluctuations 
in a minimal model for iron pnictides. 
The model consists of two bands on a square lattice exhibiting four Fermi 
pockets and  
a transition from the normal to a nematic state. It is shown that 
the orbital fluctuations produce in the $B_{1g}$ channel strong quasi-elastic 
light scattering around the nematic critical temperature $T_n$, both above and
below $T_n$. This holds for the $A_{1g}$ symmetry only below $T_n$
whereas no low-energy scattering from orbital fluctuations
is found in the $B_{2g}$ symmetry. Due to the nematic distortion the
electron pocket at the $X$-point may disappear at low temperatures.
Such a Lifshitz transition causes in the $B_{2g}$ 
spectrum a large upward shift of spectral weight in the high energy region 
whereas no effect is seen in the other symmetries.   
\end{abstract}


\maketitle

\section{Introduction}
Electronic nematic states are electronic analogues of nematic liquid crystals, 
which break only the orientational symmetry, but retain the other symmetries 
of the system. Electronic nematicity is discussed in a number of correlated 
electron systems such as quantum spin systems, \cite{penc11} two-dimensional 
electron gases, \cite{lilly99,du99} cuprate superconductors, 
\cite{kivelson03,vojta09}  
bilayer ruthenates, \cite{mackenzie12} and iron pnictides. \cite{fisher11}   
Depending on electronic degrees of freedom responsible for nematic order, 
we may distinguish between three kinds of nematicity. 
i) The {\it charge} nematicity which is obtained either by  
partial melting of charge stripes \cite{kivelson98} or by
a Pomeranchuk instability, \cite{yamase00,metzner00} 
ii) the {\it spin} nematicity which is driven by frustration between 
magnetic interactions, \cite{andreev84} and 
iii) the {\it orbital} nematicity due to orbital order caused, for example, 
by a spontaneous occupation difference between 
$d_{yz}$ and $d_{zx}$ orbitals  in $d$-electron systems. \cite{raghu09,wclee09} 

In iron pnictides, electronic nematicity is associated with the tetragonal-orthorhombic 
structural phase transition. \cite{fisher11} The structural transition 
is believed to be driven primarily by coupling to the electronic system. 
In fact, electronic resistivity exhibits a pronounced anisotropy by applying 
a uniaxial strain to the system. Moreover, angle-resolved photoemission 
spectroscopy \cite{fisher11,zhang11} revealed directly a sizable energy 
difference of $d_{yz}$ and $d_{zx}$ orbitals, 
indicating the importance of orbital nematicity. \cite{krueger09,cclee09,lv09,yanagi10a,kontani11}  
Since the structural transition occurs slightly above a spin-density-wave (SDW) 
phase, spin nematicity is also discussed as a plausible scenario \cite{fang08,xu08,fernandes12}.

Quite recently a nematic instability was observed also in magnetic torque 
measurements, \cite{kasahara12} which are very sensitive to the breaking of 
a fourfold symmetry. The observed critical temperature is much higher than 
the onset temperature of the SDW phase and extends to regions far away from 
the SDW phase. 
It seems therefore reasonable to associate such a nematic instability to 
orbital nematicity. \cite{kasahara12} 

Measurements of the anisotropy of the resistivity and magnetic torque provide 
only indirect evidence
of electronic nematicity. In the case of charge nematicity, 
it was shown theoretically \cite{yamase11} that the Raman spectroscopy in 
the $B_{1g}$ 
channel in a tetragonal system measures directly the charge nematic correlation function. 
Since the nematic instability does not break translational symmetry and thus is characterized 
by momentum zero, it is natural to believe that the Raman spectroscopy can become 
a suitable method to detect also orbital nematicity.  

In the present paper, we provide a microscopic understanding of Raman scattering 
by orbital nematic fluctuations. In Sec.~2 we introduce a minimal two-band model
for iron pnictides, 
which exhibits an orbital nematic instability at low temperatures and four 
Fermi pockets. The Raman scattering intensity is then computed 
in the lowest-order of the bosonic propagator associated with nematic 
fluctuations. 
Numerical results both in the normal and nematic states 
and their interpretation are presented in Sec.~3. 
The effect of Coulomb screening is also studied. 
Our conclusions then follow in Sec.~4. 

\section{model and formalism} 
\subsection{Nematic transition in a minimal two-band model}
Our model
Hamiltonian has the form $H = H_0 + H_1$ where the interaction part $H_1$
is given by
\begin{equation}
H_1 = \frac{g}{2} \sum_i n_{i-} n_{i-}.
\label{H1}
\end{equation}
The difference density operator $n_{i-}$ is defined by
$n_{i-} = n_{i1}-n_{i2}$ with the density operator 
$n_{i\alpha} = \sum_\sigma c^\dagger_{i\alpha\sigma} c_{i\alpha\sigma}$. 
$i$ and $\sigma$ are site and spin indices, respectively, and $\alpha=1,2$
is a band index. $g$ is a coupling constant which is considered 
as a parameter in our model. An expression for $H_0$ suitable for pnictides
is \cite{raghu08,yao09}
\begin{equation}
H_0 = \sum_{{\bf k}, \sigma, \alpha,\beta}\epsilon^{\alpha\beta}_{\bf k}
c^\dagger_{{\bf k}\alpha\sigma} c_{{\bf k}\beta\sigma},
\label{H0}
\end{equation}
with 
\bea
&&\epsilon^{11}_{\bf k} = -2t_1 \cos k_x -2t_2 \cos k_y -4t_3 \cos k_x 
\cos k_y,
\label{e11} \\
&&\epsilon^{22}_{\bf k} = -2t_2 \cos k_x -2t_1 \cos k_y -4t_3 \cos k_x 
\cos k_y, 
\label{e22}\\
&&\epsilon^{12}_{\bf k} = -4t_4 \sin k_x \sin k_y.
\label{e12}
\eea
Reasonable values for the 
hopping amplitudes are \cite{yao09} 
$t=-t_1,t_2/t=1.5,t_3/t=-1.2,t_4/t=-0.95$, which we will also use in
our calculations. 
The band indices $\alpha=1$ and $2$ originate mainly from 
the $d_{zx}$ and $d_{yz}$ orbitals, respectively.
In the following energies are always given in units of $t$.

Taking $H_1$ into account 
the Green's function matrix in band space 
is given in mean-field 
approximation by
\begin{equation}
\hat{G}({\bf k},\omega)=
{\begin{pmatrix}
\omega+i\delta-\xi_{{\bf k}1}       &     -\epsilon^{12}_{\bf k}         \\
-\epsilon^{12}_{\bf k}              &      \omega +i\delta-\xi_{{\bf k}2}  
\end{pmatrix}}^{-1}
\label{G}
\end{equation}
with 
\be
\xi_{{\bf k}1}= \epsilon^{11}_{\bf k} - \mu +gn_{-},
\label{xi1}
\ee
\be
\xi_{{\bf k}2}= \epsilon^{22}_{\bf k} - \mu - gn_{-},
\label{xi2}
\ee
and $n_{-} = \langle n_{i-} \rangle$. $\langle ...\rangle$ denotes
the expectation value, $\mu$ is the chemical potential, and
$\delta$ an infinitesimally small positive quantity. 
Carrying out the inversion in Eq.~(\ref{G}), rearranging terms
and using Pauli matrixes $\hat{\tau}_1,\hat{\tau}_2,\hat{\tau}_3$
as well as the $2 \times 2$ unit matrix $\hat{\tau}_0$
we find
\be
\hat{G}({\bf k},\omega) = g_0 \hat{\tau}_0 +g_1 \hat{\tau}_1
+g_2 \hat{\tau}_3,
\label{GG}
\ee
\bea
&&g_0 = \frac{1}{2} \left( \frac{1}{\omega-\lambda^+_{\bf k} +i \delta}
+\frac{1}{\omega-\lambda^-_{\bf k} +i \delta}\right),
\label{g0}\\
&&g_1 = \frac{1}{2} \frac{\epsilon^{12}_{\bf k}}{E_{\bf k}} 
\left(\frac{1}{\omega-\lambda^+_{\bf k} +i \delta}
-\frac{1}{\omega-\lambda^-_{\bf k} +i \delta}\right),
\label{gg1}\\
&&g_3 = \frac{1}{2} \frac{\xi^-_{\bf k}}{E_{\bf k}} 
\left(\frac{1}{\omega-\lambda^+_{\bf k} +i \delta}
-\frac{1}{\omega-\lambda^-_{\bf k} +i \delta}\right).
\label{gg3}
\eea
Here we used the abbreviations
\bea
&&\lambda^{\pm}_{\bf k} = \xi^+_{\bf k} \pm E_{\bf k},
\label{lambda} \\
&&\xi^{\pm}_{\bf k} = \frac{1}{2} (\xi_{{\bf k}1} \pm \xi_{{\bf k}2}),
\label{xi3}\\
&&E_{\bf k} = \sqrt{(\xi^-_{\bf k})^2 + (\epsilon^{12}_{\bf k})^2}.
\label{Ek}
\eea
$n_{-}$ satisfies the following nonlinear equation,
\be
n_{-} = \frac{2}{N} \sum_{\bf k} \frac{\xi^-_{\bf k}} {E_{\bf k}}
\left[f(\lambda^+_{\bf k})-f(\lambda^-_{\bf k})\right].
\label{n-}
\ee

\begin{figure}  
\vspace*{0cm}
\includegraphics[angle=0,width=6.0cm]{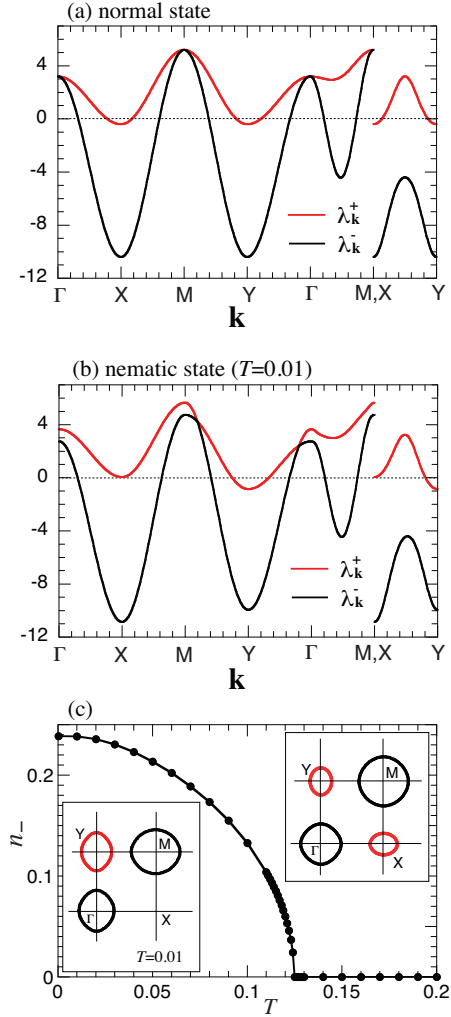} 
\caption{\label{fig1} 
(Color online) 
Dispersion of the two bands along symmetry directions in the normal
(upper panel) and nematic (middle panel) state. The Fermi energy corresponds 
to zero.  
The bottom panel shows the nematic order parameter $n_{-}$ 
as a function of temperature $T$. Left and right
insets depict the Fermi lines in the nematic state at low temperatures and
in the normal state, respectively.
}
\end{figure}

The dispersion of the two bands in the normal state is shown in the 
upper panel in \fig{fig1}. The special $\bf k$-points are 
$\Gamma = (0,0)$, $X=(\pi,0)$, $M=(\pi,\pi)$,
and $Y=(0,\pi)$, and the dispersion is drawn along the lines passing 
through these points.  
The band $\lambda^{+}_{\vk} (\lambda^{-}_{\vk})$ crosses the Fermi energy 
around $X$ and $Y$ ($\Gamma$ and $ M$) points. 
Consequently, the Fermi lines form two electron pockets at $X$ and $Y$ and 
two hole pockets at $\Gamma$ and $M$, as shown in  the right inset of the 
bottom panel. 
At lower temperature our model exhibits for negative values of $g$ a nematic phase
transition, i.e., a line which separates states where $n_{-}$ 
is zero from those where it is nonzero.
The solid line in \fig{fig1}(c) depicts the temperature
dependence of the order parameter $n_{-}$. 
In the nematic phase, where $n_{-} \neq 0$, the point group is reduced 
from $C_{4v}$ to $C_{2v}$. 
As a result the pockets at the $X$ and $Y$ points 
are no longer related by symmetry. The electron pocket 
at the $X$ point shrinks while that at the $Y$ point expands. 
When $n_{-}$ becomes 
sufficiently large a Lifshitz transition can occur at low temperatures 
where the pocket at the $X$ point vanishes. 
This is illustrated in \fig{fig1}(b) and in the left inset of \fig{fig1}(c).

\subsection{Raman scattering intensity}
The Raman scattering intensity $S^{\gamma}(\omega)$ is given by 
\begin{equation}
S^{\gamma}(\omega)= -\frac{1}{\pi} [1 + b(\omega)] {\rm Im}\chi^{\gamma}(\omega) \, ,
\label{Sw}
\end{equation}
where $b(\omega)$ is the Bose function
$(e^{\beta \omega}-1)^{-1}$ and $\beta^{-1} = T$ the temperature.
$\chi^{\gamma}(\omega)$ is the retarded Green's function 
\begin{equation}
\chi^{\gamma}(\omega) = -\frac{i}{N} \int_0^\infty dt e^{i(\omega+i{\delta})t}
\langle [\rho^{\gamma}(t),\rho^{\gamma}(0)] \rangle,
\label{chi}
\end{equation}
where $N$ is the number of lattice sites, 
$[\cdot, \cdot]$ the commutator, and $\rho^{\gamma}(t)$
the operator
\begin{equation}
\rho^{\gamma} = \sum_{{\bf k},\sigma,\alpha,\beta}\gamma_{\alpha\beta}({\bf k}) 
c^\dagger_{{\bf k}\alpha\sigma} c_{{\bf k}\beta\sigma}
\label{rho}
\end{equation}
in the Heisenberg picture. $\gamma_{\alpha\beta}({\bf k})$ is the
Raman vertex in the effective mass approximation given by 
\begin{equation}
\gamma_{\alpha\beta}({\bf k}) = \sum_{r,s} e_{r}^{i} 
\frac{\partial^2\epsilon^{\alpha\beta}_{\bf k}}{\partial k_r \partial k_s}
e^{f}_{s}. 
\label{gamma}
\end{equation}
${\bf e}^i$ and ${\bf e}^f$ are the polarization vectors of the 
incident and scattered light, respectively.

The underlying point group $C_{4v}$ gives rise to three independent
cross sections corresponding to the representations $B_{1g}, B_{2g}$,
and $A_{1g}$. The $B_{1g}$ contribution is obtained by taking the
polarization vectors ${\bf e}^i=\frac{1}{\sqrt{2}}(1,1)$ and 
${\bf e}^f=\frac{1}{\sqrt{2}}(1,-1)$ which yields
\begin{equation}
\gamma_{\alpha\beta}^{B_{1g}}({\bf k}) = \frac{1}{2}
\left(\frac{\partial^2 \epsilon^{\alpha\beta}_{\bf k}}{\partial k_x^2} - 
\frac{\partial^2
\epsilon^{\alpha\beta}_{\bf k}}{\partial k_y^2}\right).
\label{gammaB1g}
\end{equation} 
It is convenient to consider $\gamma^{B_{1g}}_{\alpha\beta}$ as a matrix 
in band space 
$\hat{\gamma}^{B_{1g}}$ and to express the dependence on $\alpha$ 
and $\beta$ in terms of Pauli matrices $\hat{\tau}_1,\hat{\tau}_2,
\hat{\tau}_3$. 
Inserting the explicit expressions
for $\epsilon^{\alpha\beta}_{\bf k}$ we obtain
\be
\hat{\gamma}^{B_{1g}} = {\gamma}^{B_{1g}}_0\hat{\tau}_0 + 
{\gamma}^{B_{1g}}_3 \hat{\tau}_3,
\label{g1}
\ee
with
\be
{\gamma}^{B_{1g}}_0 = \frac{t_1+t_2}{2}(\cos k_x - \cos k_y),
\label{g2}
\ee
\be
{\gamma}^{B_{1g}}_3 = \frac{t_1-t_2}{2} (\cos k_x + \cos k_y).
\label{g3}
\ee
In a similar way the $B_{2g}$ contribution to the Raman vertex 
is obtained by taking ${\bf e}^{i}=(1,0)$ and ${\bf e}^{f}=(0,1)$: 
\bea
&&\hat{\gamma}^{B_{2g}} = {\gamma}^{B_{2g}}_0\hat{\tau}_0 + 
{\gamma}^{B_{2g}}_1 \hat{\tau}_1,
\label{g4}\\
&&{\gamma}^{B_{2g}}_0 = -4t_3 \sin k_x \sin k_y,
\label{g5}\\
&&{\gamma}^{B_{2g}}_1 = -4 t_4 \cos k_x \cos k_y.
\label{g6}
\eea
For the polarization vectors ${\bf e}^{i}={\bf e}^{f}=\frac{1}{\sqrt{2}}(1,1)$, 
both 
$A_{1g}$ and $B_{2g}$ channels contribute to the Raman intensity.
Subtracting the latter the $A_{1g}$ contribution becomes 
\bea
&&\hat{\gamma}^{A_{1g}} = {\gamma}^{A_{1g}}_0\hat{\tau}_0 + 
{\gamma}^{A_{1g}}_1 \hat{\tau}_1 + {\gamma}^{A_{1g}}_3 \hat{\tau}_3,
\label{g7} \\
&&{\gamma}^{A_{1g}}_0 = \frac{t_1+t_2}{2} (\cos k_x + \cos k_y)
+4t_3 \cos k_x \cos k_y,
\label{g8}\\
&&{\gamma}^{A_{1g}}_1 = 4t_4 \sin k_x \sin k_y,
\label{g9}\\
&&{\gamma}^{A_{1g}}_3 = \frac{t_1-t_2}{2} ( \cos k_x - \cos k_y).
\label{g10}
\eea

\begin{figure}  
\vspace*{0cm}
\includegraphics[angle=0,width=9.0cm]{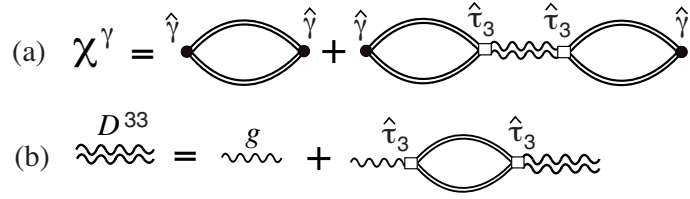} 
\caption{\label{fig2} 
Diagrams for the Raman susceptibility $\chi^\gamma$ and the propagator
$D^{33}$ for orbital fluctuations. $\hat{\gamma}$ denotes Raman
vertices and $\hat{\tau_3}$ a vertex coming from the interaction $H_{1}$. 
}
\end{figure}

Using the random phase approximation $\chi^{\gamma}(\omega)$
is given by the bubble diagrams depicted in \fig{fig2}. 
The label  $\hat{\gamma}$ stands for one of the
three Raman vertices $\hat{\gamma}^{B_{1g}}$, $\hat{\gamma}^{B_{2g}}$ or
$\hat{\gamma}^{A_{1g}}$. The vertex $\hat{\tau_{3}}$ originates from the
interaction $H_1$. The solid and wavy lines denote
the electron Green's function and the interaction $g$, respectively. 
The double wavy line represents the propagator $D^{33}$ for orbital
fluctuations. Figures~\ref{fig2}(a) and \ref{fig2}(b) are equivalent
to the following equations,
\be
\chi^\gamma(\omega) = \Pi^{\gamma \gamma}(\omega) +    
 \Pi^{\gamma 3}(\omega) D^{33}(\omega) 
\Pi^{3 \gamma}(\omega),
\label{ram}
\ee
\be
D^{33}(\omega) = \frac{g}{1-g \Pi^{33}(\omega)}.
\label{D}
\ee
Here $\Pi^{\gamma 3}$ ($\Pi^{3 3}$) denotes a bare bubble with the vertices 
$\hat{\gamma}$ and $\hat{\tau_3}$ ($\hat{\tau_3}$ and $\hat{\tau_3}$).  

Let us calculate $\Pi^{\alpha \gamma}(\omega)$
for general vertices 
\bea
&&\hat{\alpha}_{\bf k} = \alpha_0 \hat{\tau}_0 +
\alpha_1 \hat{\tau}_1 +\alpha_3 \hat{\tau}_3, 
\label{vertexa}\\
&&\hat{\gamma}_{\bf k} = \gamma_0 \hat{\tau}_0 
+ \gamma_1 \hat{\tau}_1 +\gamma_3 \hat{\tau}_3\,.  
\label{vertexg}
\eea
Performing the internal frequency sum by means of the
spectral function 
\be
\hat{A}({\bf k}, \omega) = -\frac{1}{\pi} {\rm Im} \hat{G}({\bf k},\omega),
\label{A}
\ee
and carrying out an analytic continuation we obtain for the imaginary
part of $\Pi^{\alpha \gamma}(\omega)$,
\be 
{\rm Im} \Pi^{\alpha \gamma}(\omega) = \frac{2\pi}{N} \sum_{\bf k}
\int d \epsilon {\rm Tr}\left[\hat{\alpha}_{\bf k} \hat{A}({\bf k},\epsilon)
\hat{\gamma}_{\bf k} \hat{A}({\bf k},\epsilon + \omega)\right] \left[f(\epsilon
+\omega)-f(\epsilon)\right],
\label{Pi}
\ee
where $f$ is the Fermi function. Writing
\be
\hat{A}({\bf k},\omega) = A_0 \hat{\tau}_0 + A_1 \hat{\tau}_1 
+A_3 \hat{\tau}_3,
\label{AA}
\ee
we obtain from Eqs.~(\ref{GG})-(\ref{gg3}) the following expressions for the
coefficients,
\bea
&&A_0 = \frac{1}{2\pi}\left[\frac{\delta}{(\omega-\lambda_{\bf k}^+)^2 + \delta^2}
+\frac{\delta}{(\omega-\lambda_{\bf k}^-)^2 + \delta^2}\right],
\label{A0}\\
&&A_1 = \frac{1}{2\pi}\frac{\epsilon_{\bf k}^{12}}{E_{\bf k}}
\left[\frac{\delta}{(\omega-\lambda_{\bf k}^+)^2 + \delta^2}
-\frac{\delta}{(\omega-\lambda_{\bf k}^-)^2 + \delta^2}\right],
\label{A1}\\
&&A_3 = \frac{1}{2\pi}\frac{\xi^-_{\bf k}}{E_{\bf k}}
\left[\frac{\delta}{(\omega-\lambda_{\bf k}^+)^2 + \delta^2}
-\frac{\delta}{(\omega-\lambda_{\bf k}^-)^2 + \delta^2}\right].
\label{A3}
\eea 
Carrying out the ${\rm Tr}$ in Eq.~(\ref{Pi}) the imaginary part of the considered 
bubble can be written as
\be
{\rm Im} \Pi^{\alpha \gamma}(\omega) = \frac{4 \pi}{N} \sum _{\bf k}
\int d\epsilon ({\bf G}^\alpha \cdot {\bf G}^{\gamma \prime})
\left[f(\epsilon+\omega)-f(\epsilon)\right],
\label{bubble}
\ee
where the scalar product of two four-dimensional vectors ${\bf G}^\alpha$ 
and ${\bf G}^{\gamma\prime}$ appears. 
The components of these vectors are given by 
\bea
&&G^\alpha_0 = \alpha_0 A_0 + \alpha_1 A_1 +\alpha_3 A_3,
\label{G0}\\
&&G^\alpha_1 = \alpha_0 A_1 + \alpha_1 A_0,
\label{G1}\\
&&G^\alpha_2 = -i \alpha_1 A_3 + i\alpha_3 A_1,
\label{G2}\\
&&G^\alpha_3 = \alpha_0 A_3 + \alpha_3 A_0,
\label{G3}
\eea
and
\bea
&&G^{\gamma\prime}_0 = \gamma_0 A_0^\prime + \gamma_1 A_1^\prime 
+\gamma_3 A_3^\prime,
\label{Gs0}\\
&&G^{\gamma\prime}_1 = \gamma_0 A_1^\prime + \gamma_1 A_0^\prime,
\label{Gs1}\\
&&G^{\gamma\prime}_2 = -i \gamma_1 A_3^\prime + i\gamma_3 A_1^\prime,
\label{Gs2}\\
&&G^{\gamma\prime}_3 = \gamma_0 A_3^\prime + \gamma_3 A_0^\prime.
\label{Gs3}
\eea
$A_i^\prime({\bf k},\omega)$ is defined by $A_i({\bf k},\omega+\epsilon)$. 
All the susceptibilities $\Pi$ occurring in Eqs.~(\ref{ram}) and (\ref{D})
can be obtained from the general expression Eq.~(\ref{bubble}) as special
cases. Explicit expressions are given in the appendix.  
The calculation of the various susceptibilities in the Raman scattering 
intensity $S^{\gamma}(\omega)$ [\eq{Sw}] 
has thus been reduced to one frequency and one two-dimensional momentum 
integration
which have to be carried out numerically.
The corresponding real parts can be obtained by a Kramers-Kronig 
transformation.

\section{Results}
We first present a general symmetry argument which shows that 
orbital nematic fluctuations couple in the normal state only to the 
$B_{1g}$ channel and in the nematic state to both $B_{1g}$ and $A_{1g}$ 
channels. We then present numerical results for the Raman spectra 
by computing the diagrams 
shown in \fig{fig2}. The most interesting effect due to orbital nematic 
fluctuations is the appearance of a central mode in some of the
Raman spectra. We will show that its main properties can be understood 
from analytic considerations. We will also study 
the effect of Coulomb screening on our obtained results.

\subsection{Selection rules} 
As seen in \fig{fig2}(a), orbital nematic fluctuations couple to the Raman 
susceptibility via a bubble diagram with vertices $\hat{\gamma}$ and $\hat{\tau}_{3}$. 
This diagram determines our selection rules for nematic fluctuations. 

In the normal state the symmetry group of our system is $C_{4v}$.
To see how the band basis transforms under its group elements,  
we consider each term on the right-hand side of \eq{g7}.  
In the first term it is clear that both $\gamma_{0}^{A_{1g}}$ and $\hat{\tau}_{0}$ 
have $A_{1g}$ symmetry. 
In the second term $\gamma_{1}^{A_{1g}}$ has $B_{2g}$ symmetry whereas 
the product of  $\gamma_{1}^{A_{1g}}$ and $\hat{\tau}_{1}$ should transform as  
$A_{1g}$, which means that $\hat{\tau}_{1}$ has 
$B_{2g}$ symmetry. In the third term, $\gamma_{3}^{A_{1g}}$ has $B_{1g}$ symmetry 
and thus $\hat{\tau}_{3}$ should also have $B_{1g}$ symmetry so that their 
product transforms as $A_{1g}$. 
Since $\hat{\tau}_{3}$ has $B_{1g}$ symmetry, the bubble with vertices 
$\hat{\gamma}$ and $\hat{\tau}_{3}$ is finite for $\hat{\gamma}^{B_{1g}}$ 
and vanishes otherwise. This leads to our selection rule that orbital nematic 
fluctuations can be observed in the normal state only in the $B_{1g}$ channel. 

In the nematic phase the point group is reduced to $C_{2v}$. 
$A_{1g}$ and $B_{1g}$ denote then the same representation. 
Hence not only the bubble with 
$\hat{\gamma}^{B_{1g}}$ and $\hat{\tau}_{3}$ but also that with 
$\hat{\gamma}^{A_{1g}}$ and $\hat{\tau}_{3}$ can be nonzero.  
That is, in the nematic state orbital nematic fluctuations 
can be observed in both $B_{1g}$ and $A_{1g}$ channels but not in the $B_{2g}$ 
channel.

The obtained selection rules can also be verified by 
computing $\Pi^{\gamma 3}(\omega)$ directly. In the Appendix we indeed find
in this way that 
$\Pi^{A_{1g} 3}$ is zero in the normal state and that $\Pi^{B_{2g} 3}$ is zero 
both in the normal and nematic states.

\subsection{Raman spectra} 
We compute numerically the Raman
susceptibilities \eq{ram} [see also \fig{fig2}] 
using the chemical potential $\mu = 0.6$ 
and  the coupling strength $g=-1.907$ as representative values. 
They lie in a region of the phase diagram where
the normal state at high temperatures transforms at a transition
temperature $T_n = 0.125$ into a homogenous nematic state at low 
temperatures [see \fig{fig1}(c)]. In the following calculations we choose finite
values for $\delta$ instead of taking the limit $\delta \rightarrow 0$ 
in Eqs.~(\ref{g0})-(\ref{gg3}) and (\ref{A0})-(\ref{A3}).  
This means that we replace $\delta$-functions in the spectral function
by Lorentzians with width $\delta$, which corresponds to the introduction
of a phenomenological self-energy.   
Such a procedure is necessary
because intraband contributions to the Raman scattering intensity 
can be taken properly into account only in the presence of self-energies. 
We have obtained qualitatively the same results 
for $\delta=0.05$, $0.1$, and $0.2$, 
and thus will present only the results for $\delta=0.1$. 

\subsubsection{B$_{1g}$  Raman scattering}
\begin{figure*}  
\includegraphics[angle=0,width=13.0cm]{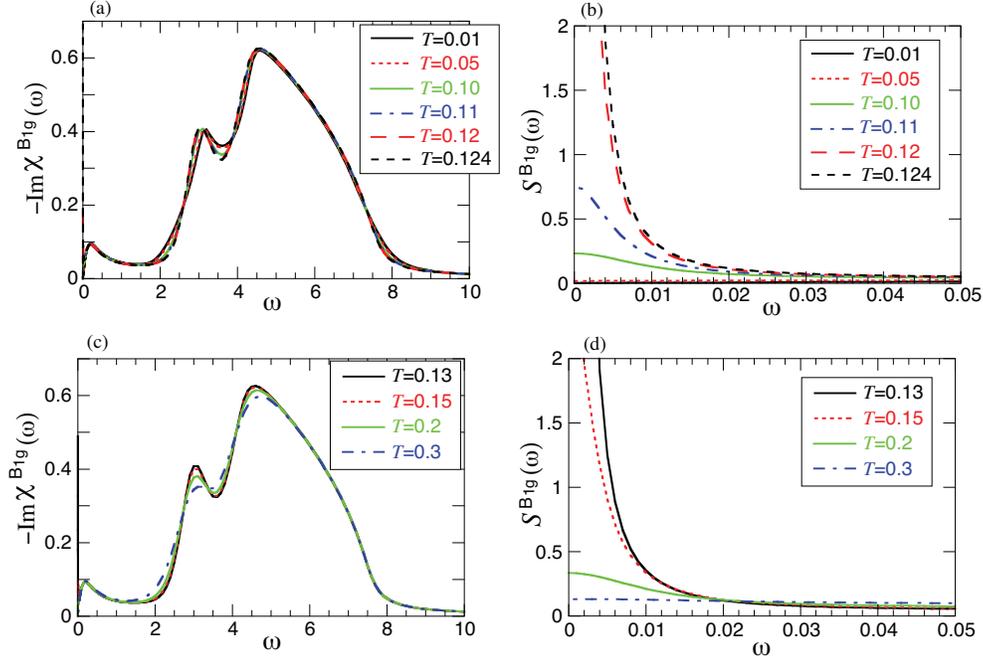} 
\caption{\label{fig3} 
(Color online) 
Panel (a) and (c): Im$\chi^{B_{1g}}(\omega)$ for temperatures
below and above $T_n$, respectively. Panel (b) and (d): $S^{B_{1g}}(\omega)$
for the same temperatures. 
}
\end{figure*}
The left panels in \fig{fig3} show Im$\chi^{B_{1g}}(\omega)$
for various temperatures below and above the transition temperature
$T_n$, respectively, over a large energy interval of the order of the 
band width. In such a representation 
no distinct temperature dependence is visible even if 
the system enters the nematic phase. 
According to Eq.~(\ref{ram}) the Raman susceptibility
$\chi^{B_{1g}}$ consists of two terms. The first one is the
bare susceptibility $\Pi^{B_{1g}{B_{1g}}}$ which is 
rather independent of temperature and dominates in the energy interval 
of the figures. 
The main peak at about $4.5 t$ arises from interband transitions
near the points $(\pi/4,\pi/4)$ and $(3\pi/4,3\pi/4)$ in agreement with the 
band structure shown in Figs.~\ref{fig1}(a) and \ref{fig1}(b).  
The effect of the second term in \eq{ram} is only moderate 
and generates a second peak at about $3t$.  
This means that in a good approximation the curves in the left panels
of \fig{fig3} represent at temperatures well below or above $T_n$
the first term in Eq.~(\ref{ram}). 

The second term in \eq{ram}, however, plays a very important role near $T_n$ 
where orbital nematic fluctuations substantially develop and become critical. 
As a result a central peak emerges in a very small frequency 
interval around $\omega=0$.  
In the right panels of \fig{fig3} we plot the scattering intensity 
$S^{B_{1g}}(\omega)$ [\eq{Sw}], 
not the imaginary part of the Raman susceptibility, to depict the spectrum 
in the very low energy region. We see that 
with decreasing temperature a central peak starts to develop below $T=0.3$, 
with a peak height at $\omega=0$ which diverges at $T=T_n$. 
Entering the nematic phase the central peak is suppressed and completely 
vanishes below $T\approx 0.05$. 
For temperatures far away from $T_n$ the total intensity is given 
approximately by the first term in \eq{ram} and is small and practically 
constant at very low energies.

\subsubsection{A$_{1g}$ Raman scattering}
\begin{figure*}  
\includegraphics[angle=0,width=13.0cm]{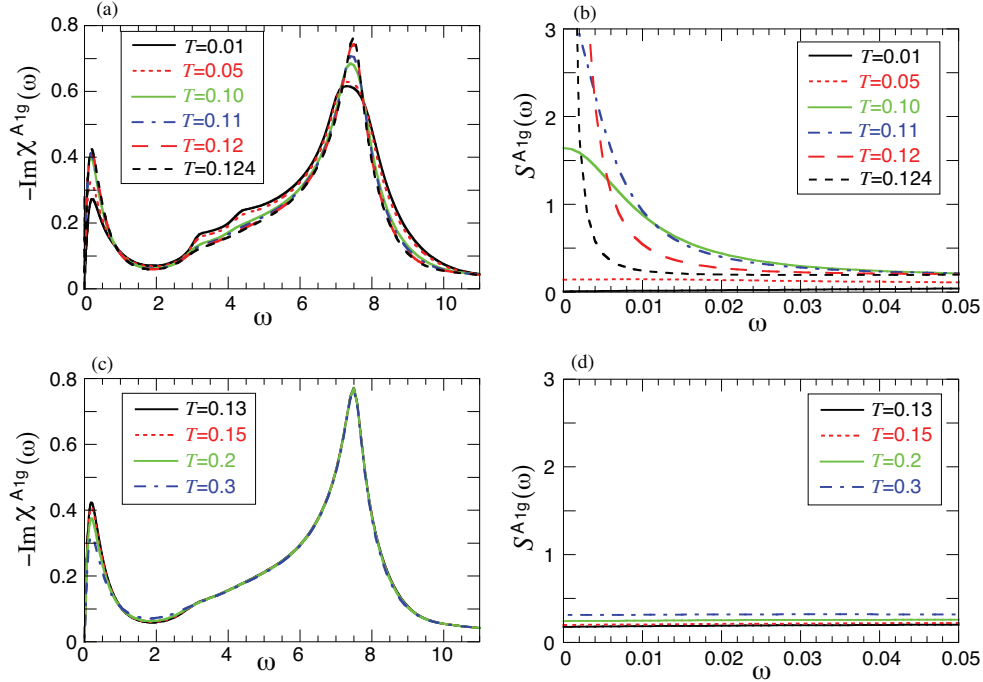} 
\caption{\label{fig4}
(Color online) 
Panel (a) and (c): Im$\chi^{A_{1g}}(\omega)$ for temperatures
below and above $T_n$, respectively. Panel (b) and (d): $S^{A_{1g}}(\omega)$
for the same temperatures. 
}
\end{figure*}
Figures~\ref{fig4}(a) and \ref{fig4}(c) show ${\rm Im}\chi^{A1g}(\omega)$ 
on a large
energy scale for temperatures below and above $T_n$, respectively. 
On this energy scale the temperature dependence of the spectra is quite weak. 
Since $\Pi^{A_{1g}3} = 0$ for $T > T_n$ (see Sec.~III~A) the curves in \fig{fig4}(c) represent
only the imaginary part of $\Pi^{A_{1g}A_{1g}}(\omega)$. They exhibit two 
well pronounced peaks. 
The main peak at $\omega \approx 7.5 t$ arises from
interband transitions near the ${\bf k}$-points $(\pi/4,3\pi/4)$ and $(3\pi/4,\pi/4)$ 
and the other peak at low energy is due to intraband transition. 
The curves in \fig{fig4}(a) include both terms in Eq.~(\ref{ram}). 
However, the contribution from the second term is minor 
in the energy interval considered in the figure and 
the curves describe essentially  $\Pi^{A_{1g}A_{1g}}(\omega)$.

Figures~\ref{fig4}(b) and \ref{fig4}(d) show the $A_{1g}$
Raman intensity in a very small frequency interval near $\omega = 0$.
In these plots the low-frequency behavior of the $A_{1g}$ Raman scattering 
becomes visible which cannot be seen in the left panels because of
their large energy scales. 
For $T>T_n$ the second term in \eq{ram} vanishes and there are no 
contributions from orbital nematic fluctuations even close to $T_n$. 
As a result the intensity is constant 
and very small at low frequencies. 
For $T<T_n$, on the other hand, orbital nematic fluctuations contribute 
substantially to the low-energy spectrum via the second term in 
Eq.~(\ref{ram}) and lead to a large central peak when $T$ 
approaches $T_n$ from below.

\subsubsection{B$_{2g}$ Raman scattering and Lifshitz transition}
\begin{figure*}  
\includegraphics[angle=0,width=13.0cm]{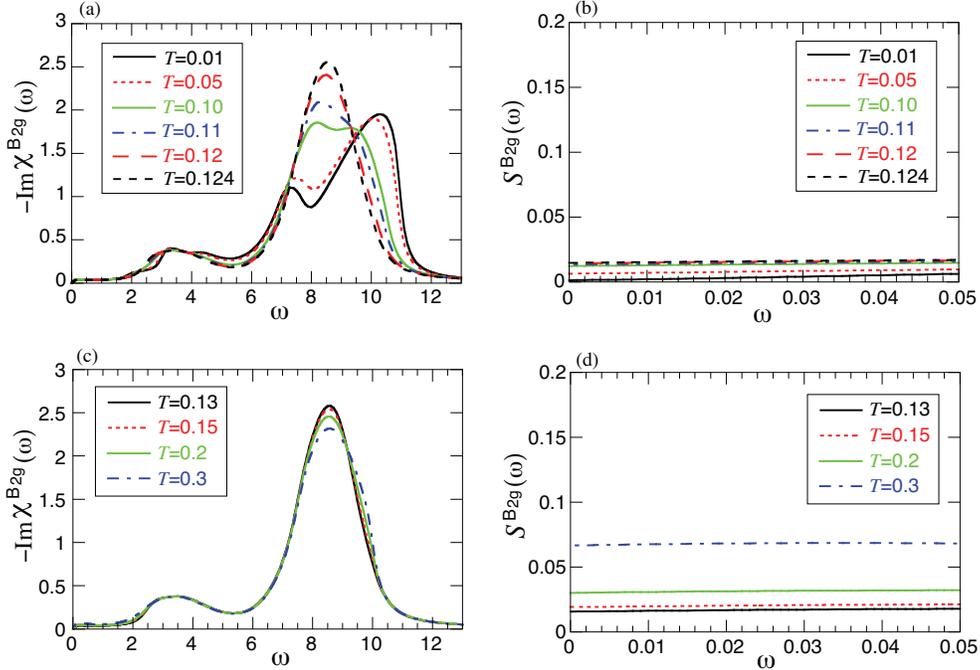} 
\caption{\label{fig5} 
(Color online)
Panel (a) and (c): Im$\chi^{B_{2g}}(\omega)$ for temperatures
below and above $T_n$, respectively. Panel (b) and (d): $S^{B_{2g}}(\omega)$
for the same temperatures. 
}
\end{figure*}
As we have discussed in Sec.~III~A orbital nematic fluctuations do not 
couple to the $B_{2g}$ component of Raman scattering which implies that the 
$B_{2g}$ spectrum is described only by $\Pi^{B_{2g} B_{2g}}(\omega)$,  
namely, by the first term in \eq{ram}.  
As a result  the Raman intensity at low frequencies becomes constant 
for all temperatures 
as shown in Figs.~\ref{fig5}(b) and \ref{fig5}(d), 
similar as for the $A_{1g}$ symmetry at temperatures above 
$T_n$ [\fig{fig4}(d)]. 
The overall increase in the intensity with increasing temperature 
in \fig{fig5}(d), especially for the $T=0.3$ curve, is simply
a result of the Bose factor in \eq{Sw}.  

Figures~\ref{fig5}(a) and \ref{fig5}(c) show the $B_{2g}$ Raman susceptibility
on a large energy scale of the order of the band width. 
For $T>T_n$ practically no temperature dependence is 
visible 
and there is a peak around $8.5t$. The peak height is by a factor of 3-4 
higher compared to 
the high-energy peak of the other symmetries  
[Figs.~\ref{fig3}(a), \ref{fig3}(c), \ref{fig4}(a), \ref{fig4}(c)]. 
With decreasing temperature and entering the nematic phase 
[\fig{fig5}(a)] the peak 
shifts to higher energies. This is a manifestation of a Lifshitz transition. 
To understand the origin for this behavior we first note that the 
peak originates 
from interband transitions which are largest at the $X$ and $Y$ points
because of the momentum dependence of the numerator in \eq{B2gB2g}.
In the normal state no interband transitions are possible at these points 
since both bands $\lambda_{\vk}^{\pm}$ are located below the Fermi energy 
[\fig{fig1}(a)]. Instead, the dominant interband transitions occur a little 
away from 
the $X$ and $Y$ points where the upper band lies above and  
the lower band below the Fermi energy. 
Considering the phase space  in the neighborhood of the $X$ point 
and the momentum dependence of the form factor in \eq{B2gB2g}, 
the peak forms near $8.5t$ in the normal state. 
However, with decreasing temperature
below $T_n$ the order parameter increases and the upper band 
at the $X$ point moves 
above the chemical potential which removes the pocket. 
Consequently strong transitions of about $10t$ are allowed near the $X$ point 
at low temperatures, which yields a peak around $10t$. 
As a result \fig{fig5}(a) exhibits a large shift of spectral weight 
towards higher energies at low temperatures. For the other symmetries 
the interband contribution to Raman scattering becomes zero at the $X$-point 
because of the momentum dependence of the form factor, which 
can be seen directly from Eqs.~(\ref{B1g3}), (\ref{A1g3}), (\ref{B1gB1g}), (\ref{A1gA1g}). 
This explains why the Lifshitz transition due to the vanishing pocket 
at the $X$-point can only be seen in the $B_{2g}$ symmetry.

\subsection{Properties of the central peak}
\begin{figure}  
\includegraphics[angle=0,width=7.0cm]{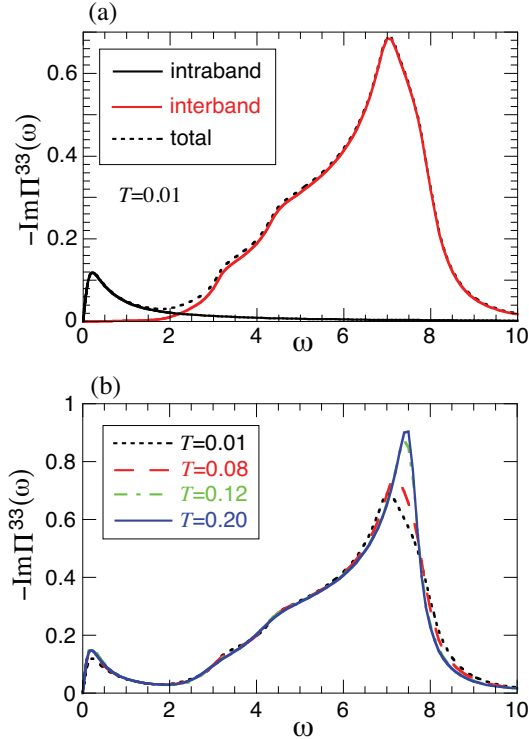} 
\caption{\label{fig6}
(Color online)
(a) Black, red, and dashed lines denote the intraband, interband and total
contribution to ${\rm Im} \Pi^{33}(\omega)$, respectively. 
(b) Im$\Pi^{33}(\omega)$ for several temperatures. 
}
\end{figure}
The central peak originates from the coupling to the nematic fluctuations 
described by \eq{D}. 
Hence the bare susceptibility for nematic fluctuations,
$\Pi^{33}(\omega)$, is an important ingredient in our calculation. 
Its explicit expression 
Eq.~(\ref{33}) shows that it may be split into two different 
contributions,
namely, an intraband contribution due to $A_{--}$ and $A_{++}$ and an
interband contribution due to $A_{-+}$ and $A_{+-}$. 
The black and red lines in the upper panel in \fig{fig6} show these two contributions for the
imaginary part of $\Pi^{33}(\omega)$, the dashed line represents
the  total ${\rm Im} \Pi^{33}(\omega)$. The figure indicates that intra- and 
interband contributions are well separated in frequency: the first one is
confined to low frequencies, increases first linearly in the frequency,
passes through a maximum near $\omega\approx 2\delta$ and then decays rapidly 
with increasing frequency. 
On the other hand, the interband contribution is very small
at small frequencies, rises with increasing frequency and shows a sharp
maximum near $\omega \approx 7t$ due to strong transitions between the
two bands near the point ${\bf k} = (\pi/2,\pi/2)$, 
see the band structure shown in \fig{fig1}(a).
The interband contribution extends over a large frequency region comparable 
to the total band width. 
The lower panel in \fig{fig6} shows
${\rm Im} \Pi^{33}(\omega)$ for several choices of temperatures. It 
depends in general only weakly on temperature
and this holds both for the intra- and the interband contributions.

The second term in Eq.~(\ref{ram})
exhibits a pole which describes in the static limit
a nematic phase transition with a transition temperature $T_n$
determined by
\be
1-g {\rm Re} \Pi^{33}(0) = 0.
\label{Tr}
\ee
It is easy to show that $T_n$ coincides 
with the largest temperature where   
Eq.~(\ref{n-}) has a non-vanishing solution for $n_{-}$, which is
the usual definition of the transition temperature. 
Since the central peak emerges near $T=T_n$ and $\omega=0$, 
we take $T-T_n$ and $\omega$ as small quantities.
Since $\Pi^{33} (\omega)$ depends only weakly on $T$ one may put
in the expression
\be
{\rm Im} D^{33}(\omega) \approx g^2 \frac{{\rm Im} \Pi^{33}(\omega)}{
\left[1-g{\rm Re}\Pi^{33}(\omega)\right]^2+
\left[g {\rm Im} \Pi^{33}(\omega)\right]^2}   
\label{ImD1}
\ee
everywhere $T=T_n$ except in the first term in the denominator.
Writing 
\be
{\rm Im} \Pi^{33}(\omega) \approx \alpha \omega
\label{alpha}
\ee
for small frequencies we obtain
\be
{\rm Im} D^{33}(\omega) \approx \frac{\omega/\alpha}{m^2(\omega)+\omega^2},
\label{ImD2}
\ee
with
\be
m^2(\omega) = \alpha^{-2}\left[{\rm Re}\Pi^{33}(0; T_n)-{\rm Re}\Pi^{33}(\omega; T)\right]^2,
\label{m}
\ee
where we denote the temperature dependence explicitly as a second argument
in $\Pi^{33}$. 

Putting in numbers one realizes that $m(0)$ is very small compared to one
for the parameter range considered by us. One reason for this is that
Re$\Pi^{33}$ depends only weakly on temperature which makes the numerator
of $m^2(\omega)$ small. As a result the second term in Eq.~(\ref{ram})
represents a low-energy contribution to the Raman susceptibility. 
Going over to the
Raman scattering intensity Eq.~(\ref{Sw}) and taking the classical limit for the
Bose function, $(1+b(\omega))/\pi \rightarrow T/(\pi\omega)$, we obtain 
approximately for
the low-energy Raman response,
\be
S^{\gamma}(\omega) \rightarrow - \frac{\left[{\rm Re} \Pi^{\gamma 3}(0; T_n)\right]^2}{\pi \alpha} \cdot
\frac{T_n}{m^2(\omega) +\omega^2}.
\label{SS}
\ee
Neglecting the $\omega$-dependence of $m^2(\omega)$ the Raman
intensity consists at low frequencies of a central peak of a Lorentzian
shape with width $m(0)$; its peak height $S^{\gamma}(0)$ is proportional to 
$1/[\alpha m^{2}(0)]$. 
The area under the central peak thus becomes proportional to 
$1/(\alpha |m(0)|)=1/|{\rm Re} \Pi^{33}(0; T_n)-{\rm Re} \Pi^{33}(0; T)|$. 
These results mean that in the limit $T\rightarrow T_n$ the width of the 
central peak 
vanishes as $|T-T_n|$ and that the peak height and 
the integrated spectral weight of the central peak diverge as $|T-T_n|^{-2}$ 
and $|T-T_n|^{-1}$, respectively. 

The asymptotic formula \eq{SS} contains the parameter $\alpha$ which 
was introduced in \eq{alpha}. $\alpha$ is determined by intraband 
scattering processes [\fig{fig6}(a)] which owe their existence to a finite
value of $\delta$ in 
Eqs.~(\ref{g0})-(\ref{gg3}) and (\ref{A0})-(\ref{A3}). In fact,  
in the limit $\delta\rightarrow 0$ $\alpha$ diverges 
and the width of the central peak vanishes even for $T\ne T_n$. 
However, the integrated spectral weight of the central peak is 
proportional to $1/|{\rm Re} \Pi^{33}(0; T_n)-{\rm Re} \Pi^{33}(0; T)|$ 
and thus independent of $\alpha$. Moreover, it is 
in general rather large because ${\rm Re} \Pi^{33}$ depends only weakly 
on temperature. 
Because of the rather weak $\delta$ dependence of Re$\Pi^{\gamma 3}(0; T_n)$, 
we thus find the remarkable 
result that the emergence of the central peak and its spectral
weight are essentially independent of the damping $\delta$.

The above analysis assumes tacitly that the linear approximation for
${\rm Im} \Pi^{33}(\omega)$ [\eq{alpha}] is valid over the region where the central
peak is substantially different from zero. The maximum of ${\rm Im} \Pi^{33}$
lies near $2\delta$ and is rather independent of $\delta$. Thus 
the linear approximation holds well in the interval $[0,2\delta]$ and
we have according to \fig{fig6} $\alpha \sim - 0.1/\delta$. Our
analysis therefore requires that $m(0) \ll 2\delta$ or that
$|{\rm Re} \Pi^{33}(0,T_n)-{\rm Re} \Pi^{33}(0,T)| \ll 0.2$, 
which is well fulfilled close to $T_n$.

\subsection{Coulomb screening} 
\begin{figure} [t]
\includegraphics[angle=0,width=7.0cm]{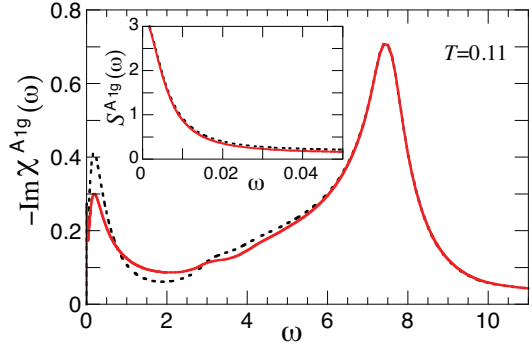} 
\caption{\label{fig7}
(Color online)
Comparison of the $A_{1g}$ spectra with (solid line) and without (dashed line)
Coulomb screening. The dashed line is
identical with the curve for $T=0.11$ in \fig{fig4}.
}
\end{figure}

It is known that the long-range Coulomb interaction may strongly
screen the first diagram in \fig{fig2}(a) in the $A_{1g}$ 
channel.\cite{klein84,monien90,devereaux07} 
This screening effect was discussed for iron pnictides
in the superconducting state and different conclusions have been obtained: In  
Refs.~\onlinecite{boyd09} and \onlinecite{mazin10} screeening was found to be important
whereas in Ref.~\onlinecite{chubukov09} it was shown that it vanishes under
plausible assumptions.
In the present section we study Coulomb screening in the normal and nematic state.   
The corresponding diagram is given by the second diagram in \fig{fig2}(a)
by replacing $\hat{\tau}_3$ by $\hat{\tau}_0$ and the coupling constant $g$
by the bare Coulomb potential $V({\bf q})$ where $\bf q$
is the momentum of the incident photon. Taking the limit
${\bf q} \rightarrow 0$ the additional contribution
due to Coulomb screening is given by
\begin{equation}
-\Pi^{\gamma 0}(\omega) \frac{1}{\Pi^{00}(\omega)} \Pi^{0 \gamma}(\omega)\,.
\label{hilf}
\end{equation}
The explicit calculation of \eq{hilf} shows that the bubbles in this expression
contain only intraband scattering processes and thus may become relevant only at low 
energy, similar as in the case of \fig{fig6}(a). Figure~\ref{fig7} shows for a representative case
$A_{1g}$ spectra with (solid line) and without (dashed line) Coulomb screening.
As expected the two curves are very similar or practical identical at energies
larger than about $t$ whereas the height of the low-energy maximum near $0.2 t$ is
noticeably reduced by screening. The inset in \fig{fig7} exhibits the 
scattering intensity $S^{A_{1g}}(\omega)$ on a very low energy scale. 
In this range $S^{A_{1g}}$ is mainly
determined by the second term in \eq{ram} which is unaffected by Coulomb screening.
This implies that the presence and the spectral form of the central mode is
insensitive to Coulomb screening. The term in \eq{hilf} enters $S^{A_{1g}}$
in the inset of \fig{fig7} only via a hardly visible reduction of the constant background.
In the normal state $\Pi^{\gamma 0}(\omega)$ is zero for $\gamma = B_{1g}$ and $B_{2g}$
so that these channels are unaffected by the Coulomb interaction in this case. 
On the other hand,  $\Pi^{\gamma 0}(\omega)$ is nonzero 
for $\gamma = B_{1g}$ in the nematic state 
(see Sec. III A). However, we find that the resulting changes in \fig{fig3} would be invisibly
small. We also have considered corrections of the nematic fluctuations due to the
Coulomb interaction, i.e., where the propagator $D^{33}(\omega)$ in \eq{D} contains
contributions from $\Pi^{00}(\omega)$ due to the non-vanishing $\Pi^{30}(\omega)$ 
in the nematic state.
We found that these effects are very small and do not change 
substantially the presented results in the nematic state. 

\section{Conclusions} 
Our analysis shows that low-energy orbital fluctuations near 
the nematic transition temperature $T_n$ can 
produce a central peak in the Raman intensity $S^{\gamma}(\omega)$. 
Depending on the symmetry
this central peak may appear both above and below $T_n$ as in the case
of the $B_{1g}$ spectrum, only below $T_n$ as for the $A_{1g}$ 
or not at all as for the $B_{2g}$ spectrum. 
The area under the central peak diverges as $|T-T_n|^{-1}$ and the 
peak width becomes narrower as $|T-T_n|$. 
While our theory contains the damping $\delta$ of the electrons, we have found 
that the integrated spectral weight of the central peak  
does not depend essentially on $\delta$. 
The predicted selection rules and 
properties of the central peak may be helpful to detect orbital
fluctuations in Raman spectra.

After the present work was completed we became aware of recent Raman scattering 
experiments in Ba(Fe$_{1-x}$Co$_x$)$_2$As$_2$ 
near the SDW phase.\cite{gallais13}  
The authors found a strong enhancement of the scattering intensity at 
low energies
around the tetragonal-orthorhombic structural phase transition 
in the $B_{1g}$ channel. Their data are consistent with our results 
[see Figs.~\ref{fig3}(b) and \ref{fig3}(d)], 
indicating that orbital nematic fluctuations become strong near the 
structural phase transition. It would be interesting to perform also 
Raman scattering measurements in the $B_{2g}$ and $A_{1g}$ channels 
in the small temperature region bounded by 
the structural phase transition and the SDW phase. 
The energy range where the enhancement of low-energy spectral weight 
was observed is, however, much wider than in our theoretical spectra 
Figs.~\ref{fig3}(b) and \ref{fig3}(d),  
if we assume $t \sim 150$ meV. 
This quantitative difference cannot be resolved by invoking a larger 
damping constant $\delta$ in 
our model. It could mean that more realistic self-energies must be included
in the calculations.

Quite recently magnetic torque measurements revealed the breaking of the  
fourfold symmetry far away from the SDW instability. As an explanation
the occurrence of an orbital nematic instability was 
discussed. \cite{kasahara12}  
It is highly desirable to perform Raman scattering measurements 
around the nematic critical temperatures measured by the magnetic torque
experiments and to confirm the presence of nematic fluctuations.  

Raman scattering in the high energy region involves mainly  
individual particle-hole excitations. At low temperatures
the nematic distortion may become large enough to induce a Lifshitz transition
where the Fermi pocket at the $X$- or the $Y$-point disappears. 
As a result particle-hole excitations at the $X$-point are allowed to occur. 
The resulting upward shift of spectral weight in the Raman intensity
should be observable
at high frequencies in the $B_{2g}$, but not in the $A_{1g}$ or $B_{1g}$ 
Raman spectra.

\begin{acknowledgments}
The authors thank D. Manske for a critical reading of 
the manuscript and M. Le Tacon and S. Tsuda for helpful discussions. 
H.Y. acknowledges support by the Alexander von Humboldt Foundation 
and a Grant-in-Aid for Scientific Research from Monkasho.       
\end{acknowledgments}
            
\bibliography{main.bib}


\begin{appendix}
\section*{Appendix}
A general formula for the bare susceptibilities has been given in
Eq.~(\ref{bubble}). By suitably specifying the components of the general 
vertices in Eqs.~(\ref{vertexa}) and (\ref{vertexg})  
one finds for each susceptibility a more explicit expression by 
computing the dot product of the four-dimensional 
vectors ${\bf G}$ in \eq{bubble}. 
It is convenient to introduce the abbreviations
\be
A_{\sigma\sigma'} = \frac{\delta}{(\epsilon+\omega-
\lambda_{\bf k}^\sigma)^2+\delta^2}
\cdot \frac{\delta}{(\epsilon - \lambda_{\bf k }^{\sigma'})^2+\delta^2},
\label{AAA}
\ee
for $\sigma=\pm, \sigma'=\pm$. The arguments 
$\epsilon,\omega,{\bf k}$ have been dropped for simplicity. 
We obtain,
\begin{eqnarray}
&&{\rm Im} \Pi^{33}(\omega) = \frac{2}{\pi N} \sum_{\bf k} \int d \epsilon
\left[f(\epsilon+\omega)-f(\epsilon)\right] \nonumber \\
&& \hspace{20mm} \times \left[ \frac{(\xi_{\bf k}^-)^2}{E_{\bf k}^2}(A_{++}+A_{--})
+\frac{(\epsilon_{\bf k}^{12})^2}{E_{\bf k}^2}(A_{+-}+A_{-+})\right];
\label{33}
\end{eqnarray}
\begin{eqnarray}
&&{\rm Im} \Pi^{B_{1g}3}(\omega) = \frac{2}{\pi N} \sum_{\bf k} \int d \epsilon
\left[f(\epsilon+\omega)-f(\epsilon)\right] \nonumber \\
&&\hspace{20mm}\times \left\{
\left[\gamma_0^{B_{1g}} \frac{\xi_{\bf k}^-}{E_{\bf k}} 
+\gamma_3^{B_{1g}} \frac{(\xi_{\bf k}^-)^2}{E_{\bf k}^2}\right]A_{++}
+\gamma_3^{B_{1g}} \frac{(\epsilon_{\bf k}^{12})^2}{E_{\bf k}^2}(A_{+-}+A_{-+})\right. \nonumber\\
&& \hspace{70mm} +\left.\left[-\gamma_0^{B_{1g}} \frac{\xi_{\bf k}^-}{E_{\bf k}} 
+\gamma_3^{B_{1g}} \frac{(\xi_{\bf k}^-)^2}{E_{\bf k}^2}\right]A_{--} 
\right\}; 
\label{B1g3}
\end{eqnarray}
\begin{eqnarray}
&&{\rm Im} \Pi^{B_{2g}3}(\omega) = \frac{2}{\pi N} \sum_{\bf k} \int d \epsilon
\left[f(\epsilon+\omega)-f(\epsilon)\right] \nonumber \\
&&\hspace{20mm}\times \left[ 
\left(\gamma_0^{B_{2g}} \frac{\xi_{\bf k}^-}{E_{\bf k}} 
+\gamma_1^{B_{2g}}\frac{\epsilon_{\bf k}^{12}\xi_{\bf k}^-}{E_{\bf k}^2}\right)A_{++}
-\gamma_1^{B_{2g}} \frac{\epsilon_{\bf k}^{12}\xi_{\bf k}^-}{E_{\bf k}^2}(A_{+-}+A_{-+}) \right.\nonumber\\
&& \hspace{70mm} +\left.\left(-\gamma_0^{B_{2g}} \frac{\xi_{\bf k}^-}{E_{\bf k}} 
+\gamma_1^{B_{2g}} \frac{\epsilon_{\bf k}^{12}\xi_{\bf k}^-}{E_{\bf k}^2}\right)A_{--}
\right]; 
\label{B2g3}
\end{eqnarray}
this susceptibility vanishes both in the normal and in the nematic state 
because the integrand contains a form factor $\sin k_x \sin k_y$ (see also Sec.~III~A 
for a symmetry-based argument); 
\begin{eqnarray}
&&{\rm Im} \Pi^{A_{1g}3}(\omega) = \frac{2}{\pi N} \sum_{\bf k} \int d \epsilon
\left[f(\epsilon+\omega)-f(\epsilon)\right] \nonumber \\
&&  \hspace{20mm} \times\left\{ 
\left[\gamma_0^{A_{1g}} \frac{\xi_{\bf k}^-}{E_{\bf k}} 
+\gamma_1^{A_{1g}} \frac{\epsilon_{\bf k}^{12}\xi_{\bf k}^-}{E_{\bf k}^2}
+\gamma_3^{A_{1g}} \frac{(\xi_{\bf k}^-)^2}{E_{\bf k}^2}\right]A_{++} \right. \nonumber \\
&& \hspace{35mm} +\left[-\gamma_1^{A_{1g}} \frac{\epsilon_{\bf k}^{12}\xi_{\bf k}^-}{E_{\bf k}^2}
+\gamma_3^{A_{1g}} \frac{(\epsilon_{\bf k}^{12})^2}{E_{\bf k}^2}\right](A_{+-}+A_{-+})  \nonumber \\
&&\hspace{50mm}+ \left. \left[-\gamma_0^{A_{1g}} \frac{\xi_{\bf k}^-}{E_{\bf k}} 
+\gamma_1^{A_{1g}} \frac{\epsilon_{\bf k}^{12}\xi_{\bf k}^-}{E_{\bf k}^2}
+\gamma_3^{A_{1g}} \frac{(\xi_{\bf k}^-)^2}{E_{\bf k}^2}\right]A_{--}
\right\};
\label{A1g3}
\end{eqnarray}
this susceptibility is zero if $n_-$ is zero, i.e., in the normal state 
because the integrand contains a form factor $\cos k_x - \cos k_y$ (see also Sec.~III~A); 
$\Pi^{3 \gamma}$ becomes equal to $\Pi^{\gamma 3}$; 
\begin{eqnarray} 
&&{\rm Im} \Pi^{B_{1g}B_{1g}}(\omega) = \frac{2}{\pi N} \sum_{\bf k} \int d \epsilon
\left[f(\epsilon+\omega)-f(\epsilon)\right] \nonumber \\
&&\hspace{20mm} \times \left[ 
\left(\gamma_0^{B_{1g}} +\gamma_3^{B_{1g}} \frac{\xi_{\bf k}^-}{E_{\bf k}}\right)^2 A_{++}
+\left(
\gamma_3^{B_{1g}} \frac{\epsilon_{\bf k}^{12}}{E_{\bf k}}
\right)^2 (A_{+-}+A_{-+}) \right.\nonumber \\
&& \hspace{70mm} + \left.\left(\gamma_0^{B_{1g}} -\gamma_3^{B_{1g}} 
\frac{\xi_{\bf k}^-}{E_{\bf k}}\right)^2 A_{--}
\right]; 
\label{B1gB1g}
\end{eqnarray}
\begin{eqnarray}
&& {\rm Im} \Pi^{B_{2g}B_{2g}}(\omega) = \frac{2}{\pi N} \sum_{\bf k} \int d \epsilon
\left[f(\epsilon+\omega)-f(\epsilon)\right] \nonumber \\
&&\hspace{20mm} \times \left[ 
\left(\gamma_0^{B_{2g}}+\gamma_1^{B_{2g}} \frac{\epsilon_{\bf k}^{12}}{E_{\bf k}}\right)^2 A_{++}
+\left(
\gamma_1^{B_{2g}} \frac{\xi_{\bf k}^{-}}{E_{\bf k}} \right)^2 (A_{+-}+A_{-+}) \right.\nonumber \\
&& \hspace{70mm}+ \left.\left(\gamma_0^{B_{2g}} -\gamma_1^{B_{2g}} \frac{\epsilon_{\bf k}^{12}}{E_{\bf k}}\right)^2 A_{--}
\right]; 
\label{B2gB2g}
\end{eqnarray}
\begin{eqnarray}
&&{\rm Im} \Pi^{A_{1g}A_{1g}}(\omega) = \frac{2}{\pi N} \sum_{\bf k} \int d \epsilon
\left[f(\epsilon+\omega)-f(\epsilon)\right] \nonumber \\
&&\hspace{10mm} \times \left\{
\left[ 
\left(\gamma_0^{A_{1g}} +\gamma_3^{A_{1g}} \frac{\xi_{\bf k}^-}{E_{\bf k}}\right)^2 
+\left(\gamma_0^{A_{1g}} +\gamma_1^{A_{1g}} \frac{\epsilon_{\bf k}^{12}}{E_{\bf k}}\right)^2
+2\gamma_1^{A_{1g}} \gamma_3^{A_{1g}} \frac{\epsilon_{\bf k}^{12}\xi_{\bf k}^-}{E_{\bf k}^2}-\left(\gamma_0^{A_{1g}}\right)^2
\right] A_{++} \right. \nonumber \\
&&\hspace{13mm} + \left. \left[ 
\left(\gamma_0^{A_{1g}} -\gamma_3^{A_{1g}} \frac{\xi_{\bf k}^-}{E_{\bf k}}\right)^2 
+\left(\gamma_0^{A_{1g}} -\gamma_1^{A_{1g}} \frac{\epsilon_{\bf k}^{12}}{E_{\bf k}}\right)^2
+2\gamma_1^{A_{1g}} \gamma_3^{A_{1g}} \frac{\epsilon_{\bf k}^{12}\xi_{\bf k}^-}{E_{\bf k}^2}-\left(\gamma_0^{A_{1g}}\right)^2
\right] A_{--} \right.\nonumber \\
&&\hspace{70mm}+\left.\left(\gamma_1^{A_{1g}} \frac{\xi_{\bf k}^-}{E_{\bf k}} - 
\gamma_3^{A_{1g}} \frac{\epsilon_{\bf k}^{12}}{E_{\bf k}}\right)^2 (A_{+-}+A_{-+}) 
\right\}. 
\label{A1gA1g}
\end{eqnarray}
\end{appendix}

\end{document}